\documentstyle[manuscript,aps,prbbib]{revtex}

\tighten
\newcommand{\be}{\begin{equation}}
\newcommand{\ee}{\end{equation}}
\newcommand{\bea}{\begin{eqnarray}}
\newcommand{\eea}{\end{eqnarray}}
\newcommand{\bd}{\begin{displaymath}}
\newcommand{\ed}{\end{displaymath}}
\newcommand{\bi}{\begin{itemize}}
\newcommand{\ei}{\end{itemize}}
\newcommand{\bc}{\begin{center}}
\newcommand{\ec}{\end{center}}
\newcommand{\bfl}{\begin{flushleft}}
\newcommand{\efl}{\end{flushleft}}
\newcommand{\bfr}{\begin{flushright}}
\newcommand{\efr}{\end{flushright}}
\newcommand{\f}{\frac}

\def\6{\partial} \def\a{\alpha} \def\b{\beta}
\def\g{\gamma} \def\d{\delta}  \def\e{\epsilon}
 \def\h{\eta} \def\th{\theta}
  \def\l{\lambda}
\def\m{\mu} \def\n{\nu} \def\x{\xi} 
\def\r{\rho} \def\ss{\sigma} \def\t{\tau}

\def\o{\omega} \def\G{\Gamma} 
\def\Th{\Theta}  \def\S{\Sigma}
  \def\O{\Omega}

\def\={\!\!\!&=&\!\!\!}
\def\+{\!\!\!&&\!\!\!+~}
\def\-{\!\!\!&&\!\!\!-~}

\newcommand{\DD}{{\cal D}}
\newcommand{\EE}{{\cal E}}
\newcommand{\FF}{{\cal F}}
\newcommand{\GG}{{\cal G}}

\newcommand{\TT}{{\cal T}}

\newcommand{\XX}{{\cal X}}

\newcommand{\ZZ}{{\cal Z}}
\def\nn{\nabla}
\def\ne{\nabla_{\eta}}
\def\nx{\nabla_{\xi}}
\def\nse{\nabla^{S}_{\eta}}
\def\nsx{\nabla^{S}_{\xi}}
\def\Rex{R(\eta ,\xi )}
\def\Rsex{R^S (\eta ,\xi )}

\begin{document}

\title{Euclidean Supergravity in Terms of Dirac Eigenvalues}
\author{Ion V. Vancea}
\address{Department of Theoretical Physics\\
Babes-Bolyai University of Cluj\\
Str. M. Kogalniceanu Nr.1, RO-3400 Cluj, Romania}
\date{\today}
\maketitle

\begin{abstract}
It has been recently shown that the eigenvalues of
the Dirac operator can be considered as dynamical variables of Euclidean
gravity. The purpose of this paper is to explore the possibility that
the eigenvalues of the Dirac operator might play the same role in the case
of supergravity. It is shown that for this purpose some primary constraints
on covariant phase space as well as secondary constraints
on the eigenspinors must be imposed. The validity of primary constraints
under covariant transport is further analyzed. It is shown that in this
case restrictions on the tangent bundle and on the spinor bundle of spacetime
arise. The form of these restrictions is determined under some simplifying
assumptions. It is also shown that manifolds with flat curvature of
tangent bundle and spinor bundle satisfy these restrictions and thus they
support the Dirac eigenvalues as global observables.
\end{abstract}

\section{Introduction}

Various attempts to understand the relationships between quantum theory
and gravity have been made by many authors along the time. The approaches
to this problem range from standard quantization methods borrowed from
quantum field theory to more sophisticated points of view upon spacetime
characterized by efforts to rethink the very structure of spacetime in 
terms of different mathematical objects other than ordinary points. These
unconventional approaches are motivated basically by the problem of 
divergences in quantum gravity which arises when one follows the standard
methods \cite{cg,msg}.

Very recently a very attractive description of quantum gravity, rather
in the framework of standard methods, was given by Landi and Rovelli 
\cite{lr}. Their results are based on some previous works by 
Chamseddine and Connes done in the framework of noncommutative geometry
\cite{cc}. Connes showed that there is a relationship between the 
geometry of a Riemannian spin manifold and the algebra generated by the 
Dirac operator together
with smooth functions on spacetime. Moreover, once the later is known, the
former can be recovered and the action of general relativity can be given
algebraically as the Dixmier trace of some function of the Dirac operator 
\cite{cc,ch,ks,mgb,kkw}. These results express the fact that the Dirac 
operator can be used
instead of the metric to describe the geometry of spacetime. Then its
eigenvalues,
which are diffeomorphism invariant objects, can be taken as dynamical 
variables of the gravitational field, which is exactly what Landi and Rovelli
did. They showed that Poisson brackets can be expressed in terms of
energy-momentum
tensor of eigenspinors which is the Jacobian matrix of transformation from
the metric to eigenvalues and that Einstein equations can be derived from a 
spectral action with no cosmological term. These very interesting results are
plagued somehow by the applicability of noncommutative geometry to only 
Euclidean case. Indeed, when one tries to extend the noncommutative geometry
to spacetime, one faces an obvious obstruction that comes from the fact
that spacetime, at macroscopic scale, has a Lorentzian structure while
the noncommutative geometry encodes the geometry of a Riemannian spin 
manifold into a real spectral triple \cite{cc}. The difference between
Lorentzian and Riemannian is precisely the obstruction here because in the
Lorentzian case one cannot give a natural positive definite inner product
for spinors on spacetime.If the positiveness is sacrified, then the Dirac
operator
is no longer self-adjoint and thus the real spectral triple is no longer
defined. Moreover, only on a Riemannian space-time manifold, 
the Dirac operator is elliptic.
However, there are tentativs to find out 
ways around this problem and an interesting
geometric construction based on a foliation of spacetime into space-like
hypersurfaces can be found in \cite{eh}. Nevertheless, the Euclidean case
is quite interesting by itself to merit further study.

It is the aim of this paper to investigate whether the eigenvalues of the 
Dirac
operator can be used to describe Euclidean supergravity. This kind of 
system has been extensively studied lately mainly in the framework of path
integral quantization of supergravity with a stress on the problem of the
boundary conditions which are to be imposed on the fermions \cite{de,ge}.
As we shall see, the extension to the minimal supergravity is possible, but
there are several constraints that must be imposed on the gravity
supermultiplet as well as on the eigenspinors of the Dirac operator. The
origin
of these constraints roots in the requirement that eigenvalues be gauge
invariant functions or dynamical variables of the system. If we require
further that the primary constraints be the same after covariant
transport along two different paths between two points, we obtain
some restrictions on the possible spacetime manifolds, more precisely
on the curvatures of tangent bundle and spinor bundle, respectively.
The form of the corresponding equations is deduced under some
siplifying assumptions and it is shown that manifolds for which
both curvatures vanish satisfy these equations. For a more general
discussion the reader is referred to a forthcoming paper
\cite{viv1} since the discussion in that case is too extensive and
presents its own distinctive problems to
be included in the present study.

The outline of the present paper is as follows. In Sec.II we review the
main results obtained in the case of general relativity. In Sec.III we
present
the construction of covariant phase space and briefly discuss the Dirac
operator
when local supersymmetry is considered. The relations that must be satisfied
by supermultiplets as well as the constraints on eigenspinors are derived
in Sec.IV. In Sec. V we determine the form of the restrictions that a
spacetime manifold should obey in order that the primary constraints mantain
their form after a covariant transport.
In Sec.VI we discuss several aspects of the theory and make
some concluding remarks.
The Appendix A reviews
some definitions from the theory of the elliptic operators necessary 
in the discussion of the Dirac operator. The Appendix B presents
the action of the two covariant derivatives used in this paper while
the Appendix C shows the most important relations necessary to
deduce the relations in Sec.V.
We use units such that $8\pi G =1$.

\section{General Relativity in Terms of Dirac Eigenvalues}

To make this paper relatively self-contained we will review the results 
obtained in the case of gravity described by Dirac eigenvalues \cite{lr}. 
We work on a compact 4D (spin) manifold without boundary $M$ and we
formulate general relativity in terms of tetrad fields $e_{\m }^{a}(x)$,
where $\m =1,\cdots ,4$ are spacetime indices and $a=1,\cdots ,4$ are
internal Euclidean indices raised and lowered by the Euclidean metric 
$\d_{ab}$. The metric field is $g_{\m \n }(x) = e_{\m }^{a}(x) 
e_{\n a}(x)$ and the spin connection 
$\stackrel{\circ}{\o_{\m b}^{a}}$ is defined by
$\6_{[\m}e^{a}_{\n ]} = \stackrel{\circ}{\o_{\m b}^{a}}e_{\n}^{b}$. 
The phase space of the 
system is covariant and is defined as the space of all solutions of the
equations of motion, modulo gauge transformations. In this case the 
gauge transformations are composed by 4D diffeomorphisms and local
rotations of the tetrad fields and thus the phase space contains equivalence
classes of tetrad fields. At the same time, the phase space can be identified
with the space of the gauge orbits on the constraint surface and with the 
space of Ricci flat 4-geometries. Let us denote the space of the smooth
tetrad fields by $\EE$ and the space of orbits of the gauge 
transformations in $\EE$ by $\GG$. The functions on the phase space are
called {\em observables} and, technically speaking, they are functions
on the constraint surface that commute with all the constraints.

On the manifold $M$ there is a natural elliptic operator, namely the Dirac
operator $\stackrel{\circ }{D}$ . The ellipticity of $\stackrel{\circ }{D}$  
means that the symbol of $\stackrel{\circ }{D}$ denoted by 
$\ss_{v}(\stackrel{\circ }{D})$ is an isomorphism (see the Appendix A). 
In terms 
of tetrads and components of spin connection the Dirac operator is given by
\be
\stackrel{\circ }{D} = i\g^a e_{a}^{\m} (\6_{\m} + 
\stackrel{\circ}{\o}_{\m bc}(e,\psi ) \ss^{bc} ), \label{doper}
\ee
where
\be
\stackrel{\circ}{\o}_{\m bc}
= \f{1}{2}e^{\n}_{b}(\6_{\m}e_{c \n}-\6_{\n }e_{c \m })
+\f{1}{2}e^{\r }_{b}e^{\ss }_{c}\6_{\ss}e_{\r d}e^{d}_{\m }
-(b \leftrightarrow c) \label{scons}
\ee
and $\g^a$'s form an
Euclidean representation of the Clifford algebra $C_4$, i. e. 
$\{ \g^a , \g^b \} = 2\d^{ab}$.  We can see from Eq.(\ref{doper}) that the 
Dirac operator is naturally defined for each geometrical structure on $M$. 
Furthermore, for each set of tetrad fields, $\stackrel{\circ }{D}$ is 
self-adjoint on the Hilbert space of spinor fields with a scalar product
\be
<\psi ,\phi > = \int d^4x \sqrt{g}\psi^{\ast}(x)\phi (x)  \label{scal}
\ee
where $\psi^{\ast}$ represents the complex conjugate of $\psi$. Since $M$ is
a compact manifold, $\stackrel{\circ }{D}$ admits a discrete spectrum of real 
eigenvalues and a complete set of eigenspinors
\be
\stackrel{\circ }{D} \stackrel{\circ}{\chi^{n}} = \stackrel{\circ}{\l^{n}}
\stackrel{\circ}{\chi^{n}} \label{spect}
\ee
where $n=0,1,2,\cdots$ . Because $\stackrel{\circ }{D}$ depends on $e$ ,
$\stackrel{\circ}{\l^n }$'s define a discrete family of real valued functions
on $\EE$ and a function from $\EE$ into the space of infinite sequences 
$R^{\infty}$
\bea
\stackrel{\circ}{\l^{n}} &:& \EE \longrightarrow R ~~~~,~~e \rightarrow
\stackrel{\circ}{\l^{n}}(e) \\ \label{funct}
\stackrel{\circ}{\l^{n}} &:& \EE \longrightarrow R^{\infty}~~,~~e \rightarrow
\{ \stackrel{\circ}{\l^{n}}(e) \}. \label{seq}
\eea
The point here is the fact that, for every $n$, $\stackrel{\circ}{\l^{n}}$
is invariant under diffeomorphisms of $M$ as well as under rotations of tetrad 
fields. Therefore they define a set of observables of general relativity.
It is worthwhile to notice that it is possible that 
$ \stackrel{\circ}{\l^{n}} $'s do not coordinate neither the space of gauge 
orbits nor the phase space. That happens any time when gauge equivalent
tetrad fields have different spectra. Then, because it is possible to find
two metric fields with the same spectra, $\stackrel{\circ}{\l^{n}} $'s do not
define an injective function.

The above construction allows us to define a Poisson structure on the set of
eigenvalues. That is possible since there is a symplectic two-form $\O$ on the
phase space given by
\be
\O (X,Y) = \f{1}{4}\int_{\S} d^3 \ss n_{\r}
[X^{a}_{\m} , \stackrel{\leftrightarrow}{\nabla_{\t}} Y^{b}_{\n} ]
\e^{\t}_{ab\upsilon } \e^{\upsilon \r \m \n} , \label{sform}
\ee
where $X^{a}_{\m}[e]$ define a vector field on the phase space and the
brackets are given by
\be
[X^{a}_{\m} , \stackrel{\leftrightarrow}{\nabla_{\t}} Y^{b}_{\n} ] =
X^{a}_{\m} \nabla_{\t} Y^{b}_{\n} - 
Y^{a}_{\m} \nabla_{\t} X^{b}_{\n} . \label{brack}
\ee
Here $\S$ is an arbitrary Arnowitt-Deser-Misner surface an $n_{\r}$ is its
normal one form. Using the inverse of the symplectic form matrix we can
write down the Poisson bracket of any two eigenvalues
\be
\{ \stackrel{\circ}{\l^{n}} , \stackrel{\circ}{\l^{m}} \} =
4\int d^4 x \int d^4 y T^{[n\m}_{a} (x) P^{ab}_{\m \n}(x,y)
T^{m ]\n}_{b} (y) , \label{pbrack}
\ee
where $ P^{ab}_{\m \n}$ is the inverse of the symplectic form matrix and
$T^{m \n}_{b}(x)$ is the energy-momentum tensor of the spinor 
$\stackrel{\circ}{\chi^{n}}$ in tetrad notation and it represents the Jacobian
matrix of the transformation from $e$ to $\stackrel{\circ}{\l^{n}}$ \cite{lr}.

As shown in \cite{ch,ks}, the gravitational action in units $\hbar 
= c =G =1$ can be written as the Dixmier 
trace of a simple function on the Dirac operator
\be
S = Tr [\g (\stackrel{\circ }{D}) ] \label{actio}
\ee
where $\g$ is a smooth monotonic function of the Dirac operator such that
\be
   \g (x) = \left\{ \begin{array}{l}
                    1 \mbox{if $x<1-\d$}  \\ 
                    0 \mbox{if $x>1+\d$} 
                    \end{array} \label{gam}
            \right. 
\ee
where $\d <<1$. Then $S$ represents the number of eigenvalues of the Dirac
operator
smaller than $1$ once the gravitino is fixed. 
One can also write the action in terms of eigenvalues
\be
S_{1} [ \stackrel{\circ}{\l} ] = \sum \g_{1}(\stackrel{\circ}{\l}) ,
\label{action}
\ee
where $ \g_{1} (x) = \g (x) - \e^4 \g (\e x)$ , $ \e << 1$ . Moreover, the
Dirac eigenvalues are not all independent and thus they cannot be simply 
varied in $S_1$ . 

There are some other interesting conclusions that can be drawn from this
variant of quantum gravity. However, because they would stray us away from
the subject, the reader is reffered to \cite{lr} for other interesting 
details.

\section{Supergravity in Terms of Dirac Eigenspinors}

To extend the ideas presented in the previous section we have to repeat
firstly the same geometrical construction in the supersymmetric case. If
we consider a supersymmetric partner of the graviton and we impose the
local supersymmetry transformations then we get Euclidean supergravity.

Consider Euclidean minimal supergravity on $M$. The graviton is represented
in the tetrad formalism by the fields $e^{a}_{\m}$. To have a local
supersymmetry we must assign to the graviton a gravitino which must be a
Majorana spinor. There is a problem here because, as is known, the group
$SO(4)$ which is the local rotation group of tetrads in the Euclidean case,
admits no Majorana spinor representation. Indeed, we can find no $SO(4)$
spinor that can satisfy Majorana condition $\psi^{\dagger}\g_4 = \psi^{T} C$.
Fortunately, there is a standard way of makeshifting around the problem. It 
is known that in the Euclidean case the following relation can be written:
$\bar{\psi} = \psi^{T} C$. Now if we define the adjoint spinors as the
ones which satisfy the previous Majorana conjugation relation we obtain the
desired Majorana spinors of the Euclidean theory. This convention does not
affect the Lorentzian theory in which the fermions are Majorana spinors. The
only difference appears in the Euclidean theory which makes now no reference
to $\psi^{\dagger}$ . With this definition of a Majorana spinor at
hand the minimal gravity supermultiplet of the theory has the right number of
degrees of freedom for both bosonic and fermionic partners \cite{ge,pn}. We
must say that, since we want to construct gauge invariant quantities, we are
interested in solutions of the equations of motion. Therefore, it is enough
to consider on-shell supersymmetry. In this case the supersymmetric algebra
closes over graviton and gravitino only. Off-shell, the supersymmetry usually
requires six more bosonic fields since there is a mismatch of the bosonic
and fermionic degrees of freedom and the supermultiplet must be enlarged
over these nonpropagating fields accordingly.

We define the phase space of Euclidean supergravity exactly as in
general relativity, namely as the space of the solutions of the equations
of motion modulo the gauge transformations \cite{cm}. The gauge
transformations are 4D diffeomorphisms, local $SO(4)$ rotations and local
$N=1$ supersymmetry. The covariant phase space is then the space of all
superpartners $(e,\psi)$ that are solutions of the equations of motion
modulo diffeomorphisms, internal rotations and local supersymmetry. As in
the nonsupersymmetric case, the observables of the theory are functions on
the phase space. Our main purpose is to see in what circumstances the
eigenfunctions of the Dirac operator can define a set of observables of 
Euclidean supergravity. To this end we must analyze the Dirac operator in the
presence of supersymmetry.

On a given spin manifold, the Dirac operator is the most fundamental differential
operator. In even dimensions the spinor space divides in half depending on
the eigenvalues of the chirality operator $\G_{D+1}$ defined as usual
\be
\G_{D+1} = \G_1 \G_2 \cdots \G_D  \label{chiop}
\ee
where $\G$'s are Dirac matrices. The chiral operator can have two real
eigenvalues and therefore any spinor enters one of the equivalence classes
defined by these two eigenvalues on the space of spinors. We can write that
as
\be
\G_{D+1} \psi_{\pm} = \pm \psi_{\pm} \label{chiopeq}
\ee
where $\pm 1$ are the two eigenvalues of the chiral operator. 
The Dirac operator
is a first order operator acting between the two chiral bundles
$C^{\infty}(S^{\pm})$. We mention that an elliptic complex can be obtained
from it by tensoring with $S^{-}$, a construction that is well known in
index
theory \cite{in}. As we mentioned in the previous section, the compactness of
$M$ ensures that the Dirac operator has a discrete spectrum and this spectrum
depends on each geometrical configuration described by tetrad fields.

Now if we consider the supersymmetric case, the local supersymmetry requires
the addition of an extra term to the Dirac operator which is of the form
\be
D= \stackrel{\circ }{D} + K \label{dir}
\ee
where $\stackrel{\circ }{D}$ is given by (\ref{doper}) and K is given by
\be
K=i\g^a e^{\m}_{a}K_{\m bc}(\psi ) \ss^{bc}  \label{bkap}
\ee
where
\be
K_{\m ab}(\psi) = \f{i}{4}(\bar{\psi_{\m}}\g_a \psi_{b} -
\bar{\psi_{\m}}\g_{b}\psi_a + \bar{\psi_b}\g_{\m}\psi_a ), \label{kap}
\ee
where
$\ss^{ab} = \frac{1}{4} [\g^{a},\g^{b}]$ and $\stackrel{\circ}{D}$ is
defined with $\ss^{ab}$, too.
If we consider that the Dirac operator is defined on the full spin bundle $SM$,
or more precisely, on the sections of it $\G (SM)$, then $D$ is an elliptic
operator. Indeed, $\stackrel{\circ }{D}$ depends only on the graviton , i. e.
on $(e,0)$ from the supermultiplet and $K$ depends only on the gravitino
$(0,\psi )$, while $D$ depends on $(e,\psi )$ which is the full gravitational
supermultiplet. Thus, the symbol $\ss_{v}(D)$ of the Dirac operator in the
presence of supersymmetry differs from $\ss_{v}(\stackrel{\circ }{D})$
by a map
\be
L'_{0} : U \rightarrow Hom(SM,SM)~~~,~~~U \subset M .
\ee
(For the definition of the symbol see Appendix A. The reader might like to
consult also \cite{asp}).
It is this map that is assigned to the term that depends on the gravitino in
(\ref{dir}). On the full spin bundle $\ss_{v}(\stackrel{\circ }{D})$ is an
isomorphism and $K(\psi )$, once the gravitino fixed, raises an isomorphism,
too.
That implies that $L'_{0}$ added to the symbol of the Dirac operator in the
nonsupersymmetric case does not affect its property of being an isomorphism.
Therefore, $\ss_{v}(D)$ is an isomorphism at its turn and from here results
that $D$ is an elliptic differential operator on $M$ in the presence of local
supersymmetry. Now if $M$ is compact as we have already assumed, $D$ has a
discrete spectrum and a complete set of eigenspinors so that we can write
\be
D \chi^n = \l^n \chi^n \label{spe}
\ee
where $n=0,1,2, \cdots$.
Let us denote the space of all gravitational supermultiplets by $\FF$. Then
$\l^n $'s define a discrete family of functions on $\FF$ since these
functions depend on $(e, \psi )$ which is a consequence of the dependence
of $D$ on $(e,\psi )$. Similar relations to (\ref{funct}), can
be written down in the supersymmetric case
\bea
\l^{n} &:& \FF \longrightarrow R ~~~~,~~(e,\psi ) \rightarrow
\l^{n}(e,\psi ) \\ \label{functs}
\l^{n} &:& \FF \longrightarrow R^{\infty}~~,~~(e,\psi ) \rightarrow
\{ \l^{n}(e,\psi ) \}. \label{seqs}
\eea
In general the eigenvalues $\l^n $'s are not invariant under the gauge
transformations of Euclidean supergravity. Therefore we cannot
immediately use $ \l^n $'s as observables. To do that we must see
under what circumstances they are gauge invariant. It is clear now that
by imposing the gauge invariance upon $\l^n$'s we must obtain some
constraints on the system. It is our purpose to find what are these
constraints. They were reported for the first time in \cite{viv}.

\section{Constraints from gauge invariance of Dirac eigenvalues
in Euclidean Supergravity}

To derive the set of all possible constraints which make the eigenvalues
of the Dirac operator gauge invariant we must impose the invariance of $\l^n$'s
with respect to each type of gauge transformation. Moreover, under the gauge
transformations Eq.(\ref{spe}) also transforms. But if we require that
$\l^n$'s
be dynamical variables, in the right-hand side of (\ref{spe}) there must
appear the  variation of eigenvalues which vanishes. The resulting equation
is an equation on the eigenspinors of the Dirac operator. Therefore, from the
gauge invariance of $\l^n$'s we must obtain constraints on the set of
eigenspinors of the Dirac operator.

Let us begin with a general variation of any eigenvalue under an 
infinitesimal gauge transformation. This is given by
\be
\d \l^n (e,\psi) = \frac{\d \l^n }{\d e^{a}_{\m}} \d e^{a}_{\m}+
\frac{\d \l^n }{\d \psi^{\a}_{\m}}\d \psi^{\a}_{\m} = 0 .
\label{diffl}
\ee
This is the fundamental relation which defines the gauge invariance of
$\l^n$. To find all of the possible constraints we must put for the
variations of the graviton and the gravitiono in (\ref{diffl}) 
their corresponding infinitesimal variations under different types of gauge
transformations.

The first type of gauge transformation is 4D diffeomorphism. This is generated
by an infinitesimal vector field on $M$ $\xi = \xi ^{\m} \6_{\m}$ where
$\xi^{\m}$ are infinitesimal. The variation of $\l^n$ under it is given by
the Lie derivatives acting on graviton and gravitino
\be
\d e^{a}_{\m} = \xi^{\n} \6_{\n} e^{a}_{\m}~~~,
~~~\d \psi_{\m} = \xi^{\n} \6_{\n} \psi_{\m} \label{diffgg}
\ee
because both $e^{a}_{\m}$ and $\psi _{\m}$ are vectors with respect to the
index $\m$. The derivative of $\l^n$ with respect to the graviton which
enters the first term in (\ref{diffl}) can be computed by deriving the scalar
product $<\chi^n , D\chi^n >$ with respect to $e^{a}_{\m}$. If we consider
that all eigenvectors are normalized we obtain
\be
\frac{\d \l^n }{\d e^{a}_{\m}}=
<\chi^n | \frac{\d }{\d e^{a}_{\m}} D | \chi^n>= 
\TT^{n \m}_{a} (x).
\label{diffd}
\ee
In deriving (\ref{diffd}) we took into account the fact that $K_{\m ab}$ 
does not
depend on the graviton. 

Here $\TT^{n \m }_{a} = T^{n\m }_{a} + K^{n\m }_{a}$
where
$T^{n \m}_{a}$ is the energy-momentum tensor  of
the spinor $\chi^n$ \cite{lr} and 
$K^{n\m }_{a}= <\chi^n |i\g_{a}K^{\m }_{bc}(\psi ) \ss^{bc}|\chi^n >$.
In a similar manner one can evaluate
the derivative of $\l^n$ with respect to the gravitino field. This is given
by the derivative of the same scalar product as in (\ref{diffd}) with respect
to the gravitino. The only term contributing to this derivative is $K$ and we
obtain after simple calculations
\be
\f{\d \l^n}{\d \psi^{\a}_{\m}} = \frac{i}{4}\int \sqrt{e}
{\chi^n}^{\ast}\g^a e_{a}^{\n}
[\bar{\psi_{\n}^{\b}}(\g_{b})_{\a \b }e^{\m}_{c} -
\bar{\psi_{\n}^{\b}}(\g_{c})_{\a \b }e^{\m}_{b} +
\bar{\psi_{b}^{\b}}(\g_{\n})_{\a \b }e^{\m}_{c}]\ss^{bc}  \chi^{n} 
= \G^{n \m }_{\a}. \label{gamm}
\ee
Eq.(\ref{gamm}) represents nothing else but the matrix elements of $K$ on
the eigenstates of $D$. If we put together (\ref{gamm}),
(\ref{diffgg}) and (\ref{diffl}) and if we consider that the variation of
$\l^n$ must vanish for arbitrary $\e^{\n}$ we obtain the following equation:
\be
\TT^{n\m }_{a} \6_{\n}e^{a}_{\m} - \G^{n \m }_{\a}\6_{\n}\psi_{\m}^{\a} =0. 
\label{firstc}
\ee
This is a first set of constraints that must be imposed on the supermultiplet
$(e,\psi)$. In a similar manner the invariance of the eigenvalues under
$SO(4)$ leads to new constraints. The fields $e$ and $\psi$ transform under an
infinitesimal $ SO(4)$ rotation as a vector and a spinor, respectively. The
corresponding relations are
\bea
\d e^{a}_{\m} &=& \th^{ab} e_{b\m} \\
\d \psi_{\m}^{\a} &=&  \th^{ab} (\ss_{ab})^{\a}_{\b} \psi_{\m}^{\b} 
\label{rot}
\eea
where $\th_{ab}=-\th_{ba}$ parametrize an infinitesimal rotation
and $\ss^{ab} = i\S^{ab}$. The
infinitesimal transformation of an eigenvalue is given by the basic relation
(\ref{diffl}) where now (\ref{rot}) must be taken into account for the
infinitesimal variation of the supermultiplet while the derivatives of $\l^n$
with respect to $e$ and $\psi$ remain the same as above. Then the following
equation is straightforward
\be
\TT^{n\m }_{a}e_{b\m} +\G^{n\m }\ss_{ab}\psi_{\m}=0. 
\label{secondc}
\ee
Eq.(\ref{secondc}) form the second set of constraints that must be imposed on
the phase space of the theory. It is equally easy to derive the last set
of constraints on $(e, \psi )$. They come from the invariance of the Dirac
eigenvalues under the local $N=1$ supersymmetry. Under an infinitesimal
on-shell supersymmetry transformation the supermultiplet transforms as
follows
\be
\d e^{a}_{\m} = \frac{1}{2}\bar{\e}\g^{a}\psi_{\m}~~~~~
\d \psi_{\m} = \DD_{\m}\e                \label{susy}
\ee
where $\e (x)$ is an infinitesimal Majorana spinor field, i. e. it obeys
the Majorana conjugation relation $\bar{\e}=\e^{T}C$.  Here $\DD_{\m}$ is the
non-minimal covariant derivative acting on spinors.
There is another minimal covariant derivative which acts on tensors and which
is expressed in terms of Christoffel symbols
(see Appendix B).
Under (\ref{susy}) the spin connection transforms as
\be
\d \o_{\m}^{ab} = A_{\m}^{ab} - \frac{1}{2}e_{\m}^{b}A_{c}^{ac}
+\frac{1}{2}e_{\m}^{a}A_{c}^{bc} \label{trsc}
\ee
where
\be
A_{a}^{\m \n} = \bar{\e}\g_{5}\g_{a}\DD_{\l}\psi_{\r}\e^{\n \m \l \r}. 
\label{aterm}
\ee
Now to derive the constraints imposed by the local supersymmetry we start as
in the diffeomorphism case and the rotation cases from the (\ref{diffl}) and we
take for the variations of $e$ and $\psi$ (\ref{susy}). Then we obtain
the following equation:
\be
\TT^{n\m }_{a}\bar{\e}\g^{a}\psi_{\m} + \G^{n\m }\DD_{\m}\e =0. \label{thirdc}
\ee
which is the expression of the constraints that must be imposed on the
phase space if $\l^n$ are invariant under $N=1$ local supersymmetry.

Let us examine now what are the consequences of the gauge invariance of
Dirac eigenvalues upon the Dirac eigenspinors. If we start with the eigenvalue
problem (\ref{spe}) and transform it under an infinitesimal gauge
transformation its variation reads
\be
\d D\chi^n = (\l^n - D)\d \chi^n \label{varr}
\label{vard}
\ee
where we considered that $\l^n$'s are invariant under an infinitesimal
gauge transformation. If the gauge transformation is a diffeomorphism of
$M$ generated by an infinitesimal vector field $\e = \e^{\n}\6_{\n}$ we have
the variations in (\ref{varr})
\be
\d D = {\pounds}_{\e} D = [\e ,D]~~~,~~~\d \chi^n ={\pounds}_{\e}\chi^n =
\e^{\n}\6_{\n} \chi^n  .\label{varf}
\ee
We use in (\ref{varf}) the expression of $D$ given by (\ref{dir}). Then a
short and simple algebraic calculus gives us the variation of $D$.
Using some short-hand notations for the terms entering this variation we
obtain
\be
\d D = i\g^a [b^{\m}_{a}(\e)\6_{\m} + f_{a}(\e)]=
[b^{\m}(\e)\6_{\m} +f(\e)] \label{varfs}
\ee
where we have used the following notations:
\bea
b^{\m}(\x )  &=& i \g^{a} b_{a}^{\m}(\x )~~,~~  
b_{a}^{\m}(\x ) = \x^{\n}\6_{\n}e_{a}^{\m} -e_{a}^{\n}\6_{\n}\x^{\m}
-2e_{a}^{\n}\x^{\m}\o_{\n bc}\ss^{bc}\\  
f(\x ) = i\g^{a}\x^{\n}\6_{\n}(e_{a}^{\m}\o_{\m bc})\ss^{bc}. \label{bfc}
\eea
Then from (\ref{varf}) and (\ref{varfs}) we obtain the following equation:
\be
\{ [b^{\m}(\x ) - c(\l ,\x )^{\m} ]\6_{\m} + f(\x ) \} \chi^n = 0
\label{consp}
\ee
where
\be
c(\l ,\x )^{\m} = (\l^n - D)\x^{\m} \label{ccc}
\ee
arises from the variation of $\chi^n$. Eq.(\ref{consp}) represents a first
set of equations that must be satisfied by the eigenspinors of the 
Dirac operator
as a consequence of the invariance of eigenvalues under diffeomorphisms.
In the case of an $SO(4)$ rotation the transformation of $D$ is slightly more
complicated since now $\o_{\m ab}$ transform as gauge fields. In this case
$e^{\m}_{a}$ transforms as a vector and $\chi^n$ transforms as a spinor.
The transformation of $\o_{\m ab}$ under rotations is given by
\be
\d \o_{\m ab} = i [{\bf \th \ss}, \o_{\m ab} ] -
i\6_{\m}{\bf \th \ss}M_{ab} \label{spinc}
\ee
where $\th_{ab} = - \th_{ba}$ parametrize an infinitesimal $SO(4)$ rotation
and ${\bf \th \ss} = \th_{ab}\ss_{ab}$. The variations of $D$ and $\chi^n$
can be obtained after some simple algebra and they are
\bea
\d D & = & \th^{a}_{a} D - \g^{a}e_{a}^{\m}
\{ [ \th \ss , \o_{\m cd} ] - \6_{\m}\th \ss M_{cd} \}
\ss^{cd} \\
\d \chi^{n \a } &=& i (\th_{ab}\ss_{ab})^{\a}_{\b} \chi^{n\b} .
\label{rottr}
\eea
Now if we introduce the following notations:
\bea
g(\th ) &=& [\g^c e_{c}^{\m}([{\bf \th \ss},\o_{\m ab}] -
\6_{\m}{\bf \th \ss }M_{ab})]\ss^{ab} \\
h(\th ) &=&  i (\l^n - D) {\bf \th \ss} \label{notrot}
\eea
and introduce (\ref{rottr}) in (\ref{vard}) the constraints coming from
$SO(4)$ invariance can be written as
\be
[\th_{a}^{a}D -g(\th ) +h(\th ) ]\chi^n = 0. \label{consts}
\ee
Finally, if $N=1$ local supersymmetry is considered, the left-hand side of
($\ref{vard}$) vanishes since $\chi^n$ are inert under this symmetry. Thus, 
our equation becomes $\d D \chi^n =0$. In this case $e^{a}_{\m}$ and
$\psi_{\m}$ tranform accordingly to (\ref{susy}). Now considering (\ref{susy})
and (\ref{trsc}) we can easily write down the variation of $D$ which leads us
immediately to the supersymmetric constraint. The calculus raising no problem
and being quite simple we write down only the result in a notation designed
to make it more transparent
\be
[j^{\m}_{a} (\e ) \6_{\m} + k_{a} (\e ) +l_{a} ]\chi^n =0 \label{contsu}
\ee
where
\bea
j^{\m }_{a} (\e ) & = & \frac{1}{2}\g_{a}\bar{\e}\psi^{\m} ~,
~~~k_{a}(\e ) = \frac{1}{2}\g_{a}\bar{\e}\psi^{\m}\o_{\m cd}\ss^{cd} \\
l_a &=&e_{a}^{\m}[B_{\m cd} - \frac{1}{2}e_{\m d}B_{ec}^{e} +
\frac{1}{2}e_{\m c}B_{ed}^{e}]\ss^{cd}.
\eea
and $\bar{\e}$ is an infinitesimal Majorana spinor. The equations 
(\ref{firstc}), (\ref{secondc}) and (\ref{thirdc}) represent constraints
on the phase space which must be imposed in order to have Dirac eigenvalues 
as observables of the theory. In this respect, they are primary constraints,
i.e. they come first into discussion when the dynamical variables are 
discussed. As consequences of these primaries follow the equations 
(\ref{consp}), (\ref{consts}) and (\ref{contsu}) which restrict the set of
eigenspinors to those which satisfy these relations. It is fair to say that 
all these equations are highly nontrivial. A method to solve them is unknown
to the author at present. It would be interesting to see whether the
constraint equations have any solution because there is no obvious evidence
that this is the case. For example, even if the conditions of gauge invariance
of Dirac eigenvalues are fulfilled, it might be possible that there be no
eigenspinor to satisfy the constraints on spinors. That, in turn, would mean
that there are no eigenspinors compatible to this description of Euclidean
supergravity which would question the validity of this approach. The lesson
to be learnt from here is that both sets of equations must have nontrivial
solutions in order to have a consistent formulation of Euclidean
supergravity in terms of Dirac eigenspinors.

Let us point out some subtleties which would appear in a possible
quantization of Euclidean supergravity in terms of Dirac eigenvalues. As
in the gravity case, the information about the geometry of the manifold
in the presence of local supersymmetry is encoded in $\l^n$'s. At a first
glance, in the process of quantization the eigenspinors could be left aside
since they do not appear directly and we should pay attention only to the
eigenvalues. This conclusion is not true. As we saw in Sec. II, the Poisson
brackets of $\l^n$'s are determined by energy-momentum tensor of the
eigenspinors and thus the eigenspinors become important in the quantization
process. In the supersymmetric case the situation is somehow similar, but
the fact that the system is subjected to several constraints makes it more
suitable for BRST quantization \cite{then}. The eigenspinors intervene in
quantization precisely through these constraints. To see how this comes, let
us denote the constraints (\ref{firstc}), (\ref{secondc}) and (\ref{thirdc})
with $\S^n (\TT, \G ) =0$, $\Th^{n}_{ab} (\TT, \G) =0$ and 
$\Phi^n (\TT, \G ) =0$.
Here $\TT$ and $\G$ denote $T^{n\m }_{a}$ and $\G^{n\m }_{a}$, respectively. The
path integral can be written using the Fadeev-Popov trick as
\be
\ZZ \sim \int D[e]D[\psi ]e^{- S_0 } \sim
\int D[e]D[\psi ]D[\ss ]D[\t ]D[\phi ] e^{-S} \label{partf}
\ee
where the first integral is factorized with the volume of all of the gauge
transformations and
\bea
S & = & S_0 +S_1 +S_2  \\
S_1 &=& \int (\ss_n \S^n (\TT,\G ) +\t_{n}^{ab} \Th^{n}_{ab} (\TT, \G ) +
\phi_n \Phi^n (\TT, \G )) \\
S_2 &=& \int (s_n \d_{\a} \S^n (\TT,\G ) +
t_{n}^{ab} \d_{\a} \Th^{n}_{ab} (\TT, \G ) +
f_n \d_{\a} \Phi^n (\TT, \G ))c^{\a} \label{tact}
\eea
where $\ss_n $,$\t_{n}^{ab}$ and $\phi_n$ are the antighosts associated to
the gauge averaging conditions, $s_n$,$t_{n}^{ab}$ and $f_n$ are the
corresponding ghosts and $c^{\a}$ are the ghosts associated to the gauge
transformations and denoted generically by $\d_{\a}$. We observe that the
quantities $\TT^{n\m }_{a}$ and $\G^{n\m }_{\a}$ enter the path integral. But
they are computed as matrix elements between eigenspinors of the Dirac 
operator
and these eigenspinors must be solutions of (\ref{consp}), (\ref{consts}) and
(\ref{contsu}). This implies that one should take only those matrix elements
of $\TT$ and $\G $ that are obtained in those eigenstates of the Dirac operator
which also solve the above constraints. Obviously, this leads to a certain
simplification of the path integral. But the price to be paid for this is
solving the constraints on the eigenspinors, which in turn implies solving
the primary constraints. To illustrate this point we could have employed a
more complete formulation of the BRST theory, like BV or BFV, but the 
problems
remain the same. Since our discussion has the role to emphase the 
difficulties
which arise in the quantization of Euclidean supergravity, we paid no
attention to the structure of gauge transformation which is vital for the
quantum theory. However, as long as the constraints of the theory are not
solved, any further discussion of the quantization has just a general
character. The matter deserves a deeper study, but that is out of the line
of the present paper.

\section{Global consistency of primary constraints}

Primary constraints (\ref{firstc}), (\ref{secondc}) and (\ref{thirdc})
have a rather local character. However, to obtain a theory fully
consistent, it would be desirable that these constraints be compatible
with the global structure of $M$. As usual, when a global problem is
addressed, this compatibility might restrict the possible manifolds that
can support the theory. If secondary constraints are promoted to global
constraints, too, further restrictions may appear. However, since
secondaries arise as consequences of primaries and since they essentially
restrict only the spinors belonging to the Dirac eigenspinors set,
we do not investigate this matter here.

The main idea which will be used in what follows to study the consequences of
globality of primary constraints, is to transport these equations from one
point of $M$ to another one along two different paths. If we require that
the same result be obtained after the two transports we find some
restrictions upon the structure of $M$.

The transport of any of the constraints will be implemented via the
exponentiation of covariant derivative, similar to the exponentiation
of Lie derivative\footnote{For spinors there is difficult to define
Lie derivative along an arbitrary vector field, but if this vector
field is chosen to have a particular form, for example to be a
conformal Killing vector, the problem is removed.}. Since the constraints are
made out of objects which are composed by bosons and fermions,
such as $\TT^{n\m }_{a}$ or $\G^{n\m}_{\a}$, we must consider a covariant
derivative acting on bosons and another one acting on fermions. These
are associated to two connections $\nabla$ and $\nabla^S$ on the tangent
bundle $TM$ and spinor bundle $SM$ respectively
\bea
\nabla &:& \XX (M) \rightarrow Hom_{R} (TM,TM) \nonumber \\
\nabla^S &:& \XX (M) \rightarrow Hom_R (SM,SM) ,
\eea
where $\XX (M)$ represents the algebra of vector fields on $M$. Due to the
composite structure of constraints, the natural way to transport them from
one point to another is to transport firstly all of the elementary
objects as graviton, gravitino and the derivative, and then to reconstruct
more complicated objects as the components of spin connection or of $K$ term
and then to write down the whole equation.

Let us assume that we have two congruences $c(\l )$ and $d(\m )$ on $M$, where
$\l$ and $\m$ are the parameters of the curves, and let us select a
curvilinear rectangle at the intersection of the two congruences
\be
\{ Q, P, R, S, \} \in c(\l ) \cap d(\m )
\ee
where $|QP|\in c$, $|RS| \in c$, $|PR| \in d $, $|QS|\in d$ and in this case
$|~~|$ denotes the curvilinear segment. Let us assume that the lengths of
the sides of the rectangle are $\l$ and $\m$ measured in units of natural
parameters of $c(\l )$ and $d(\m )$, respectively. Suppose further that there
are two vector fields $\x $ and $\h $ from $\XX (M)$ such that $\x $ is
defined along $c(\l )$ and $\h$ is defined along $d(\m )$ and
\be
[\x ,\h ]=0. \label{commvf}
\ee
As a particular case we can take $\x = d/d{\l}$ and $\h = d/d{\m}$.

Now any object $A$ can be transported, say, from $Q$ to $P$ along $c(\l)$ and
the result will be
\be
A(P) = e^{\l \bar{\nabla}_{\x}}A(Q) \label{transporta}
\ee
where $\bar{\nabla}$ stands for either $\nabla$ or $\nabla^S$ and
\be
\bar{\nabla}_{\x} A(Q) = \x^{\n}\DD_{\n} A(Q)   \label{deriva}
\ee
where $\DD_{\n}$ is the covariant derivative. It acts on bosons minimally
and on fermions non-minimally acccording to the requirements of
supersymmetry ( see Appendix B). The minimal covariant derivative is related
to $\nabla$ and the nonminimal one to $\nabla^S$.

In what follows we are interested in transporting the constraints along
$Q\rightarrow P\rightarrow R$ which we call path 1 and along
$Q\rightarrow S\rightarrow R$ which we call path 2. Any object carrying
the subscript 1 or 2 will be understood as transported along the
respective path. Thus, for example, for a boson transported along path 1
we have
\be
A_1 = e^{\m \nabla_{\h}}e^{\l \nabla_{\x}}A . \label{transpoa}
\ee
An equality between two objects transported along path 1 or path 2 should
hold at all orders in power expansion of exponentials in (\ref{transpoa}).
We expect that the coefficients of $\m \l $ capture some information about
the structure of $M$. To make this term important we take $\m$ and $\l$
be conveniently small and thus the series in (\ref{transpoa}) truncate to
\be
A_1 = A^{(0)}_1 + \m A^{(1)}_1 + \l A^{(2)}_1 + \f{\m^2}{2} A^{(3)}_1
    + \f{\l^2}{2} A^{(4)}_1 + \m \l A^{(5)} . \label{sertra}
\ee

Before moving to the transport of constraints, let us make some general
remarks that will ease the forthcoming calculus otherwise pretty heavy
and long. Any of the equations (\ref{firstc}), (\ref{secondc}) and
(\ref{thirdc}) consists of a sum of two terms each of these being written
as a product $AB$. Now when we compare the transported equations along the
two paths we come to sums of two terms of the form $A_1 B_1 - A_2 B_2 $
where $A_1 $ and $B_1 $ can be cast into the form (\ref{sertra}). If we remark
that under the change of paths 1$\leftrightarrow$2 (\ref{sertra}) displays
the following symmetry:
\bea
A^{(0)}_1\leftrightarrow A^{(0)}_2~~
&,&~~A^{(3)}_1 \leftrightarrow A^{(4)}_2 \nonumber\\
A^{(1)}_1 \leftrightarrow A^{(2)}_2~~
&,&~~A^{(4)}_1 \leftrightarrow A^{(3)}_2 \nonumber\\
A^{(2)}_1 \leftrightarrow A^{(1)}_2 ~~
&,&~~A^{(5)}_1 \leftrightarrow \bar{A}^{(5)}_2  ,\label{symma}
\eea
where the change of paths implies in fact
\be
\m \leftrightarrow \l ~~,~~\h \leftrightarrow \x \label{symmatoo}
\ee
and where we have considered that $\bar{A}^{(5)}$ is the same as
$A^{(5)}_2$ but with the product $\nabla_{\h} \nabla_{\x}$ inverted.
We must also notice that for any $A$, $A^{(0)}_1 = A^{(0)}_2 = A^0 $.
Taking (\ref{symma}) and (\ref{symmatoo}) into account we see that
\be
A_1 B_1 - A_2 B_2 = A^0 (B^{(5)}_{1NS} -B^{(5)}_{2NS})+
(A^{(5)}_{1NS} -A^{(5)}_{2NS})B^0 , \label{funda}
\ee
where the index $NS$ indicates the non-symmetric part in $\nabla_{\h}$
and $\nabla_{\x}$. The same holds true for fermions with the appropriate
connection. The symmetry above can be easily demonstrated using the
following relation:
\be
e^{\m \nn_{\h}} e^{\l \nn_{\x}} = 1+\m \nn_{\h} +\l \nn_{\x}
+\f{\m^2}{2}\nn_{\h}\nn_{\h} +\f{\l^2}{2}\nn_{\x}\nn_{\x} +
\m \l \nn_{\h}\nn_{\x} \label{fundab}
\ee
which also holds for the spinorial case, with $\nn$ replaced
by $\nn^S$.

Another important point that should be discussed concerns the Dirac
eigenspinors $\chi^n $. In general, an arbitrary spinor changes while it is
transported along an arbitrary path. In the case of Dirac eigenspinors, this
change is unwanted since it can take an eigenspinor $\chi^n$ out of the set of
eigenspinors or it can move it onto another eigenspinor corresponding to
a different eigenvalue $\chi^m$. Both these changes alter our quantities
$\TT^{n\m }_{a}$,$\G^{n\m }_{\a }$ and therefore we must require that they
be zero. To this end, if we transport the Dirac equation along path 1 we
must have
\be
D_1 \chi^{n}_{1} = \l^{n}_{1} \chi^{n}_{1} \label{diractr}
\ee
where $D_1$ is given by (\ref{dir}) with all of the
entering objects transported along path 1. Spin connection is transported by
transporting each graviton
in it, according to (\ref{scons}). Analogously, for the $K$ be transported we
should firstly transport each fermionic component. The results are given in
Appendix C. Using them we can write down the transported Dirac equation and
we can power expand it. Acoording to the previous discusion we consider only
the equivalent of (\ref{funda}) for this case. We are eventually led to
\bea
(D-\l^n )(\Rsex \chi^n ) &=& \{ \Rex \l^n -i\g^{a} [(\Rex e^{a}_{\m })(\6_{\m}
+\f{1}{2}\stackrel{\circ}{\o_{\m bc}\ss^{bc}}) \nonumber \\
&+& i\g^a e^{a}_{\m }(\Rex \6_{\m }+
\f{1}{2}\o^{(5)}_{\m bc}(\Rex ,\Rsex )\ss^{bc})] \}\chi^n  ,\label{diractt}
\eea
where $\Rex$ and $\Rsex$ are the curvatures of the tangent bundle $TM$ and
of the spinor bundle $SM$, respectively. $\o^{(5)}_{\m bc}(\Rex ,\Rsex )$
can be obtained from $\stackrel{\circ}{\o}^{(5)}_{\m bc}(\Rex ) $ and
$K^{(5)}_{\m bc}(\Rsex )$ from Appendix C. The dependence of 
$\o^{(5)}_{\m bc}(\Rex ,\Rsex )$ on $\Rex$ and $\Rsex$ means that in
$\o^{(5)}_{\m bc}(\Rex ,\Rsex )$ the products $\ne \nx$ and $\nse \nsx$
must be replaced by $\Rex$ and $\Rsex$, respectively.
Indeed, once that (\ref{commvf}) holds
true, the two curvatures are given by
\be
\Rex = [ \ne ,\nx ] ~~~,~~~ \Rsex = [ \nse , \nsx ]. \label{curvatures}
\ee

Equation (\ref{diractt}) can be viewed as a restriction of possible
curvatures of tangent bundle and spinor bundle once the eigenspinor
$\chi^n $ is given. In principle, it should hold for all spectra
of the
Dirac operator. Thus we see that this equation imposes restrictions on
the structure of $M$, namely on the two curvatures mentioned above. 
These restrictions are consequences of the promotion of
the eigenspinors to global observables of Euclidean supergravity. 
The presence of the two curvatures in this equation is exactly
the type of restriction we should have expected from this kind of treatment
of the problem of global consistency of constraints. Notice that before
discussing the transport of the Dirac equation along the two paths considered
above we should have discussed a simple transport, i.e. the transport between
just two arbitrary points. This would have led us to constraints on the
two connections instead of constraints on the two curvatures. 
To see that, let us assume that we transport the Dirac
equation only along $c(\l )$. The equation must hold true at the final point.
Expanding the exponential in powers of $\l $ we obtain an infinte set of
equations. The first two of them are the following ones:
\be
D\chi^n =\l^n \chi^n  \label{diracsecc}
\ee
at order zero and
\be
[ D ( \nsx \chi^n )+i\gamma^a (\nx e^{\m}_{a})(\6_{\m} + 
\f{1}{2}\stackrel{\circ}{\o_{\m bc}}\ss^{bc})
+ i\gamma^a e^{\m}_{a}((\nx \6_{\m})+ \f{1}{2}\o^{\x (0)}_{\m bc} \ss^{bc} )] \chi^n
= (\nx \l^n )\chi^n + \l^n (\nsx \chi^n ) \label{diracsec}
\ee
at first order.
Here $\o^{ \x (0)}_{ \m bc} $ can be obtained from $\o^{(0)}_{ \m bc}$ if we
set $exp(\m \ne ) $ to one. The former equation is automatically satisfied
while the second one remains as a constraint. The rest of constraints play
an important or less important role depending on the magnitude of the
parameter $\l $. Working in the real global case, that is when we move
on an arbitrary distance on the curve $c(\l )$, the power expansion is
less useful and we have to work with the exponentials instead. For a more
detalied analysis of these issues we relegate the reader to the
forthcoming paper \cite{viv1}.

An obvious simplification of (\ref{diractt}), (\ref{diracsecc})
and (\ref{diracsec}) can be obtained if the eigenspinors are subject to 
parallel transport
along $c(\l )$ and $d(\m )$
\be
\nse \chi^n = \nsx \chi^n = 0. \label{parallelhi}
\ee
As a consequence, once $\chi^n$ is fixed in a point on $M$ it remains the same
as a spinor field. We shall assume in what follows that that is the
case. A more general discussion can be found in \cite{viv1}.

Let us analyze what happens when the primary constraint (\ref{firstc})
is transported along the two paths. To compute the changes in
$\TT^{n \m}_{a}$, $\G^{n \m}_{\a}$, $\6_{\n }e^{a}_{\m }$ and
$\6_{\n}\psi^{\a}_{\m}$ we use (\ref{fundab}) and the Appendix C. After
some tedious algebra the final results can be expanded in powers of
$\l$ and $\m$ as in (\ref{sertra}). As shown in (\ref{funda}), the essential
terms are the coefficient of order zero and of $\m \l$ non-symmetric in
$\ne\nx$ and $\nse\nsx$ (see Appendix C). Using them as well as
(\ref{curvatures}) and (\ref{parallelhi}) we obtain after some 
algebra the following relation:
\bea
<\chi^n |i \g^d \d_{da} \d^{\m \r}
(\6_{\r} + \f{1}{2}\stackrel{\circ}{\o}^{(0)}_{\r fg}\ss^{fg}) -
\nonumber\\
\f{1}{8}\g_a g^{\m \n} \sum_{(\n ,b,c)}
[\bar{\psi}_{\n} \g_b \psi_c ]
\ss^{bc}|\chi^n >
[(\Rex )\6_{\n})e^{a}_{\m} +
\6_{\n}(\Rex e^{a}_{\m})] +
\nonumber\\
<\chi^n | i \g^d \d_{da} \d^{\m \n}
[(\Rex \6_{\n} ) +
\f{1}{2}\stackrel{\circ}{\o}^{(5)}_{\n fgNS}(\Rex )\ss^{fg}]
+
\nonumber\\
i\g^d[\f{\d}{\d e^{a}_{\m}}(\Rex e^{\r}_d )
(\6_{\r} + \f{i}{2}\stackrel{\circ}{\o}^{(0)}_{\r fg}\ss^{fg}) -
\f{\d}{\d e^{a}_{\m}}[(\Rex e_{\n d}] \d^{\n \r}
(\6_{\r} + \f{1}{2}\stackrel{\circ}{\o}^{(0)}_{\r fg}\ss^{fg})] -
\nonumber\\
\f{1}{8}\g_a g^{\m \r}\sum_{(\r ,b,c)}
[\bar{\psi}_{\r}\g_b (\Rsex \psi_c ) + (\Rsex \bar{\psi_{\r}})\g_b \psi_c ]
\ss^{bc}|\chi^n> \6_{\n} e^{a}_{\m} +
\nonumber\\
\f{1}{8}<\chi^n | \g^a e^{\r}_{a} \f{\d }{\d \psi^{\a}_{\m}} 
[\sum_{(\r ,b,c)}
[\bar{\psi}_{\r}\g_b \psi_c ]]\ss^{bc}|\chi^n>
[(\Rsex \6_{\n} )\psi^{\a}_{\m} +
\6_{\n}(\Rsex \psi^{\a}_{\m})]-
\nonumber\\
<\chi^n |i \g^a \{ e^{\r}_{a} \f{\d}{\d \psi^{\a}_{\m}}
[\f{i}{8}\sum_{(\r ,b,c)}[(\Rsex \bar{\psi}_\r )\g_b \psi_c +
\bar{\psi}_\r \g_b (\Rsex \psi_c )] +
[\d^{\m}_{\n} \d^{\b }_{\a } (\Rex e^{\r}_{a})-
\nonumber\\
\f{1}{8}\f{\d}{\d \psi^{\a }_{\m }} 
( \Rsex \psi^{\b }_{\n }) e^{\r }_{a}] 
\f{\d}{\d \psi^{\b}_{\n}}[\sum_{(\r ,b,c)}[\bar{\psi}_{\r}\g_b \psi_c ]]
\} \ss^{bc} |\chi^n > \6_{\n} \psi^{\a}_{\m} = 0
\label{onetransp}
\eea

We can work out the second primary constraint (\ref{secondc}) along the
same line. Performing exactly the same steps we obtain the second
restriction on the manifold $M$. Using again the results listed in
the Appendix C we find the following equation:
\bea
<\chi^n |i \g^d \d_{da} \d^{\m \n}
(\6_{\n} + \f{1}{2}\stackrel{\circ}{\o}^{(0)}_{\n fg}\ss^{fg}) -
\nonumber\\
\f{1}{8} \g_a g^{\m \n} \sum_{(\n ,d,c)}
[\bar{\psi}_{\n} \g_d \psi_c ]
\ss^{dc}|\chi^n > 
(\Rex e_{b \m}) +
<\chi^n | i \g^d \d_{da} \d^{\m \n}
[(\Rex \6_{\n} ) +
\nonumber\\
\\f{1}{2}\stackrel{\circ}{\o}^{(5)}_{\n fgNS}(\Rex )\ss^{fg}]
+i\g^d[\f{\d}{\d e^{a}_{\m}}(\Rex e^{\r}_d )
(\6_{\r} + \f{1}{2}\stackrel{\circ}{\o}^{(0)}_{\r fg}\ss^{fg}) -
\nonumber\\
\f{\d}{\d e^{a}_{\m}}[(\Rex e_{\n d}] \d^{\n \r}
(\6_{\r} + \f{1}{2}\stackrel{\circ}{\o}^{(0)}_{\r fg}\ss^{fg})] -
\nonumber\\
\f{1}{8}\g_a g^{\m \n}\sum_{(\n ,d,c)}
[\bar{\psi}_{\n}\g_d (\Rsex \psi_c ) + (\Rsex \bar{\psi_{\n}})\g_d \psi_c ]
\ss^{dc}|\chi^n> e_{b \m} +
\nonumber\\
<\chi^n | \g^d e^{\r}_{d} \f{\d }{\d \psi^{\a}_{\m}} 
\f{1}{2}[\sum_{(\r ,t,c)}
[\bar{\psi}_{\r}\g_t \psi_c ]]\ss^{tc}|\chi^n>
[\ss_{ab}]^{\a}_{\g }(\Rsex \psi^{\g}_{\m}) 
+
\nonumber\\
<\chi^n |i\g^d \{ e^{\r}_{d} \f{\d}{\d \psi^{\a}_{\m}}
[\f{1}{2}\sum_{(\r ,t,c)}[(\Rsex \bar{\psi}_\r )\g_t \psi_c +
\nonumber\\
\bar{\psi}_\r \g_t (\Rsex \psi_c )] +
[\d^{\m}_{\n} \d^{\b }_{\a } (\Rex e^{\r}_{d})+
\nonumber\\
\f{1}{2}\f{\d}{\d \psi^{\a }_{\m }} 
( \Rsex \psi^{\b }_{\n }) e^{\r }_{d}] 
\f{\d}{\d \psi^{\b}_{\n}}[\sum_{(\r ,t,c)}[\bar{\psi}_{\r}\g_t \psi_c ]]
\} \ss^{tc} |\chi^n > [\ss_{ab}]^{\a}_{\g}\psi^{\g}_{\m} = 0
\label{twotransp}
\eea
In a similar manner we obtain the final restriction on the manifold from the
third primary (\ref{thirdc}). The terms that multiply $\TT^{n \m}_{a}$ and
$\G^{\n \m}_{\a}$ are now $\bar{\e}\g^a \psi_{\m} $ and $\DD_{\m}\e^{\a}$.
The transport of the value of covariant derivative of $\e^{\a}$
from $Q$ to $R$ along of the path 1 is performed in a similar manner to
the transport of the Dirac operator which led us to (\ref{diractt}).
The calculations of the restriction imposed by the third primary constraint
are somewhat more lenghty than the previous computations and heavily rely on
the Appendix C, too. The final result is
\bea
<\chi^n |i \g^d \d_{da} \d^{\m \n}
(\6_{\n} + \f{1}{2}\stackrel{\circ}{\o}^{(0)}_{\n fg}\ss^{fg}) -
\nonumber\\
\f{1}{8} \g_a g^{\m \n} \sum_{(\n ,b,c)}
[\bar{\psi}_{\n} \g_b \psi_c ]
\ss^{bc}|\chi^n > 
[(\Rsex \bar{\e})\g^a \psi_{\m} +
\bar{\e}\g^a (\Rsex \psi_{\m})] +
\nonumber\\
<\chi^n | i \g^d \d_{da} \d^{\m \n}
[(\Rex \6_{\n} ) +
\f{1}{2}\stackrel{\circ}{\o}^{(5)}_{\n fgNS}(\Rex )\ss^{fg}]
\nonumber\\
+i\g^d[\f{\d}{\d e^{a}_{\m}}(\Rex e^{\r}_d )
(\6_{\r} +\f{1}{2} \stackrel{\circ}{\o}^{(0)}_{\r fg}\ss^{fg}) -
\f{\d}{\d e^{a}_{\m}}[(\Rex e_{\n d}] \d^{\n \r}
(\6_{\r} + \f{1}{2}\stackrel{\circ}{\o}^{(0)}_{\r fg}\ss^{fg})] -
\nonumber\\
\f{1}{8}\g_a g^{\m \n}\sum_{(\n ,b,c)}
[\bar{\psi}_{\n}\g_b (\Rsex \psi_c ) + (\Rsex \bar{\psi_{\n}})\g_b \psi_c ]
\ss^{bc}|\chi^n> \bar{\e}\g^a \psi_{\m} +
\nonumber\\
\f{1}{8}<\chi^n |i \g^a e^{\r}_{a} \f{\d }{\d \psi^{\a}_{\m}} 
[\sum_{(\r ,b,c)}
[\bar{\psi}_{\r}\g_b \psi_c ]]\ss^{bc}|\chi^n>
[(\Rex \6_{\m} )\e^{\a} +
\nonumber\\
\6_{\m}(\Rsex \e^{\a})+
\f{1}{2}[\stackrel{\circ}{\o}^{(5)}_{\m abNS}(\Rex )[\ss^{ab}]^{\a}_{\b}
\e^{\b}+
\nonumber
\stackrel{\circ}{\o}^{(0)}_{\m ab}[\ss^{ab}]^{\a}_{\b}
(\Rsex \e^{\b})]+ 
\nonumber\\
\f{1}{8}[\sum_{(\m ,b,c)}[\bar{\psi}_{\m}\g_b \psi_c ][\ss^{bc}]^{\a}_{\b}
(\Rsex \e^{b}) + \sum_{(\m ,b,c)}[\bar{\psi}_{\m}\g_b (\Rsex \psi_c )+
\nonumber\\
(\Rsex \bar{\psi}_{\m})\g_b \psi_c ][\ss^{bc}]^{\a}_{\b}\e^{\b}]+
<\chi^n |i\g^a \{ e^{\r}_{a} \f{\d}{\d \psi^{\a}_{\m}}
[\f{1}{8}\sum_{(\r ,b,c)}[(\Rsex \bar{\psi}_\r )\g_b \psi_c +
\nonumber\\
\bar{\psi}_\r \g_b (\Rsex \psi_c )] +
[\d^{\m}_{\n} \d^{\b }_{\a } (\Rex e^{\r}_{a})+
\nonumber\\
\f{\d}{\d \psi^{\a }_{\m }} 
( \Rsex \psi^{\b }_{\n }) e^{\r }_{a}] \times
\nonumber\\
\f{\d}{\d \psi^{\b}_{\n}}[\f{1}{8}
\sum_{(\r ,b,c)}[\bar{\psi}_{\r}\g_b \psi_c ]]
\} \ss^{bc} |\chi^n > \DD_{\m}\e^{\a} = 0
\label{threetransp}
\eea

Equations (\ref{onetransp}), (\ref{twotransp}) and (\ref{threetransp})
can be interpreted as restrictions over the possible spacetime
manifolds. Specifically, they restrict the curvatures of the tangent
bundle $TM$ and of the spinor bundle $SM$. We note that these restrictions
have been obtained for the case of two congruences on which two commuting 
vector
fields $\x$ and $\eta$. Another severe assumption made in deducing
the restrictions above was that the Dirac eigenspinors undergo 
parallel transport
along the two vector fields. The most general cases are analyzed in
\cite{viv1}.

The restrictions have not a nice form, even though they display some symmetry. 
Since they have such a complicate structure it is difficult to say
how the curvatures should look for the manifold to permit, under the
above hypothesis, the covariant transport of primary constraints on it.
Nevertheless, the system of restrictions posses a trivial solution
\be
\Rex = \Rsex = 0 \label{trivial}
\ee
which assures that at least manifolds with flat tangent bundle and also flat 
spinor bundle admits Dirac eigenspinors as global obsrvables.

\section{Discussions and Concluding Remarks}

Throughout this paper we analyzed under what circumstances the eigenvalues of
the Dirac operator can be used as observables of Euclidean supergravity. We 
saw
that, as in the case of general relativity, the Dirac operator in the presence 
of
local supersymmetry is an elliptic differential operator of first order. Then,
on a compact (spin) manifold, it admits a discrete spectrum. However, its
eigenspinors might not coordinate the covariant phase space of the theory
which is composed by equivalence classes of gravitational supermultiplets
through equations of motion. This might happen because the eigenvalues might
fail to define bijective functions from the geometry to the set of infinite
sequences, as is the case in nonsupersymmetric theory. If we insist 
that the theory
should be described in terms of Dirac eigenvalues, we come to the crucial 
difference
with respect to general relativity: the eigenvalues are no longer gauge 
invariant by
construction. The gauge transformations of the present theory are spacetime
diffeomorphisms, local rotations and $N=1$ local supersymmetry. From the
invariance of Dirac eigenvalues under gauge transformations, which is
necessary in order to have them playing the role of dynamical variables, we
obtain the equations (\ref{firstc}), (\ref{secondc}) and (\ref{thirdc}).
These equations represent manifest constraints on the phase space. From them
follow a second set of relations, namely (\ref{consp}), (\ref{consts}) and
(\ref{contsu}) which represent additional restrictions on the eigenspinors of
the Dirac operator. To have a consistent theory, both sets of constraints must
have nontrivial solutions. We also saw that these constraints play a
major role not only in the classical theory, but also in the quantization
process. Next we addressed the problem of global extension of these
observables over the manifold $M$. To investigate this matter, we consider
transporting the primary constraints from a given point to another one,
the two being at the opposite corners of a rectangle made by four
points that belong to the intersection of two congruences. Subject to
some simplifying assumptions, we concluded that the curvature of the
tangent bundle and that of the spinor bundle must satisfy the equations
(\ref{onetransp}), (\ref{twotransp} and (\ref{threetransp}). Even if these
restrictions are expressed by rather complicated equations, it was shown that
manifolds with flat tangent bundle and flat spinor bundle satisfy them.
Therefore, on these manifolds, the Dirac eigenvalues can be promoted to
global observables in the sense mentioned above. This is not surprisingly
since flat manifolds seem to be related to supersymmetric theories.

\acknowledgements
I would like to thank P. Blaga for discussions and J. Gates Jr. for useful
suggestions.

\section*{Appendix A}

The symbol of an operator is an useful object in the study of the properties
of differential operators defined on a manifold. These operators allow us
to get relevant information about the manifold. These constructions are most
frequently met in physics in the context of gauge theories where elliptic
complexes and Atiyah-Singer index theorem are tools in use\cite{in}.

Let us locally define the symbol of a differential operator \cite{tp}. 
For the beginning, if $A$
and $B$ are vector spaces over $R$($C$), $U \subset R^n $ and $E= U
\times A$, $F=U \times B$, then a {\em differential operator} of order $k$
is an $R$($C$)-linear operator $D: \G (E) \rightarrow \G (F)$ such that
for each $n$-tuple $\a = (\a_1 , \cdots , \a_n )$ of nonnegative integers
there exists an application 
$L_{\a} :U \rightarrow Hom(A, B)$ such that for all
$f \subset \G (E)$
\be
D(f) = \sum_{ |\a | \leq k }L_{\a}D^{\a} f, \label{dop}
\ee
where
\be
D^{\a} = \f{\6^{ | \a | }}{\6 x_{1}^{\a_1 } \cdots \6 x_{n}^{\a_n } }
~~~,~~~| \a | = \sum_{i =1}^{n} \a_i.
\ee
Here $\G (E)$ is the set of sections of $E$. Now let
$y = (y_1 , \cdots , y_n ) \in R^n $ and
$y^{\a} = y^{\a_1 }_{1} \cdots y^{\a_n }_{n}$ and $v=(x,y)\in U \times R^n $.
Then there is an application $A\rightarrow B$ defined as follows
\be
\ss_v (D) = \sum_{|\a |} y^{\a }L_{\a } (x)
\ee
which defines a map from $U \times R^n \rightarrow Hom(A,B)$ called the
{\em symbol} of $D$, denoted by $\ss (D)$ and given by $v \rightarrow
\ss_v (D)$. We say that $D$ is {\em elliptic} if $\ss (D)$ is an isomorphism
for all $v = (x,y)$ with $y \neq 0$.

The extension of these definitions to vector bundles is straightforward
\cite{tp}. We note only that there exists a symbol of $D$ defined as follows:
for $x\in M$, $v\in T^{*}_{x}M $, $\ss_v (D): E_x \rightarrow F_x $ where
$E_x $ and $F_x$ are the fibres over $x$ of two vector-bundles $E$,$F$ over
the same smooth manifold $M$ for which $\pi :T^* M \rightarrow M$ is the
projection of the tangent bundle. Then for $e \in E_x $ and $s \in \G (E)$
with $s(x) = e$ and $g \in C^{\infty} (M)$ with $g(x)=0$ and $dg(x)=v$
the symbol of $D: \G (E) \rightarrow \G (F)$ is given by
\be
\ss_{v} (D)(e) = D (\f{g^k }{k!} s) (x) \in F_x  .
\ee
The map $v \rightarrow \ss_v (D)$ defines an element
$\ss (D) \in \G Hom(\pi^* E,\pi^* F )$ where $E_x \simeq (\pi^* E)_v ,
F_x \simeq (\pi^* F )_v $ .

\section*{Appendix B}

In this appendix we present the action of the two covariant derivatives on
different objects. The minimal covariant derivative acting on a world vector
$A=A^{\m }\6_{\m }$ is given by
\be
\nx A^{\n} = 
\x^{\m}(\6_{\m}A^{\n} + \G^{\n}_{\m \ss}A^{\ss})=
\x^{\m}\DD_{\m}A^{\n}
\label{minder}
\ee
and the covariant derivative of $SO(4)$ vectors $A=A^{a }\6_{a}$ is given
by
\be
\nx A^{a} = 
\x^{\m }(\6_{\m}A^{a} + \o_{\m b}^{a}A^{b})
=\x^{\m }\DD_{\m } A^a .
\label{mindero}
\ee
The covariant derivatives of the gravity supermultiplet are given by
\be
\nx e^{a}_{\m }
= \x^{\n}(\6_{\n} e^{a}_{\m } - \G^{\r}_{\m \n}e^{a}_{\r}
+ \o^{a}_{\n b}(e,\psi )
e^{b}_{\m } )= \x^{\n}\DD_{\n}e^{a}_{\m }
\label{spindero}
\ee
for the graviton and
\be
\nsx \psi^{\a}_{\n} =
\x^{\m}(\6_{\m} \psi^{\a}_{\n} +\frac{1}{2}\o_{\m ab}(e,\psi ) \ss^{ab}
\psi^{\a}_{\n})
=\x^{\m}\DD_{\m} \psi^{\a}_{\n}
\label{spinder}
\ee
for gravitino.
The "supersymmetric spin connection" contains the usual spin connection and
$K_{\m ab}$ term depending on the gravitino
\be
\o_{\m ab}(e,\psi ) =  \stackrel{\circ}{\o_{\m ab}}(e) + K_{\m ab}(\psi ).   
\label{conn}
\ee

\section*{Appendix C}

Here are given some useful objects transported along path1:
\bea
e^{a}_{\m 1} &=& e^{\m \ne}e^{\l \nx} e^{a}_{\m }
\nonumber\\
\psi^{\a}_{\m 1} &=& e^{\m \ne}e^{\l \nx} \psi^{\a}_{\m }
\nonumber\\
\6_{\m 1} &=& e^{\m \ne}e^{\l \nx} \6_{\m }
\nonumber\\
{\f{\d}{\d e^{a}_{\m}}}_1 &=&
\f{\d (e^{-\l \nx}e^{-\m \ne}e^{b}_{\n})}{\d e^{a}_{\m}}
{\f{\d}{\d e^{b}_{\n}}}
\nonumber\\
{\f{\d}{\d \psi^{\a}_{\m}}}_1 &=&
\f{\d (e^{-\l \nx}e^{-\m \ne}\psi^{\b}_{\n})}{\d \psi^{\a}_{\m}}
{\f{\d}{\d \psi^{\b}_{\n}}}
\eea
The essential coefficients of power expanding of the components of
the primary constraints are given by
\bea
\TT^{n\m (0)}_{a 1} =
<\chi^n |i \g^d \d_{da} \d^{\m \n}
(\6_{\n} + \f{1}{2}\stackrel{\circ}{\o}^{(0)}_{\n fg}\ss^{fg}) +
\f{i}{8}\g^d \d_{da} g^{\m \n} \sum_{(\n ,b,c)}
[\bar{\psi}_{\n} \g_b \psi_c ]
\ss^{bc}|\chi^n > 
\nonumber\\
\TT^{n\m (5)}_{a1NS} =
<\chi^n | i \g^d \d_{da} \d^{\m \n}
[(\ne \nx \6_{\n} ) +
\f{1}{2}\stackrel{\circ}{\o}^{(5)}_{\n fgNS}(\ne \nx )\ss^{fg}]+
\nonumber\\
i\g^d[\f{\d}{\d e^{a}_{\m}}(\ne \nx e^{\r}_d )
(\6_{\r} +\f{1}{2}\stackrel{\circ}{\o}^{(0)}_{\r fg}\ss^{fg}) +
\f{\d}{\d e^{a}_{\m}}[(\nx \ne e_{\n d}] \d^{\n \r}
(\6_{\r}+
\nonumber\\
\f{1}{2}\stackrel{\circ}{\o}^{(0)}_{\r fg}\ss^{fg})] +
\f{i}{8}\g^d \d_{da} g^{\m \n}\sum_{(\n ,b,c)}
[\bar{\psi}_{\n}\g_b (\nse\nsx \psi_c ) +
(\nse\nsx \bar{\psi_{\n}})\g_b \psi_c ]
\ss^{bc}|\chi^n> 
\nonumber\\
\G^{n\m (0)}_{\a 1} =
<\chi^n |\f{i}{8} \g^a e^{\r}_{a} \f{\d }{\d \psi^{\a}_{\m}} 
[\sum_{(\r ,b,c)}
[\bar{\psi}_{\r}\g_b \psi_c ]]\ss^{bc}|\chi^n>
\nonumber\\
\G^{n\m (5)}_{\a1NS} =
<\chi^n |i\g^a \{ e^{\r}_{a} \f{\d}{\d \psi^{\a}_{\m}}
\f{1}{8}[\sum_{(\r ,b,c)}[(\nse\nsx \bar{\psi}_\r )\g_b \psi_c +
\bar{\psi}_\r \g_b (\nse\nsx \psi_c )] +
[\d^{\m}_{\n} \d^{\b }_{\a } (\ne\nx e^{\r}_{a})+
\nonumber\\
\f{\d}{\d \psi^{\a }_{\m }} 
(\nsx\nse \psi^{\b }_{\n }) e^{\r }_{a}] 
\f{\d}{\d \psi^{\b}_{\n}}[
\f{1}{8}\sum_{(\r ,b,c)}[\bar{\psi}_{\r}\g_b \psi_c ]]
\} \ss^{bc} |\chi^n > 
\nonumber\\
\6_{\n}e^{a (0)}_{\m 1} = \6_{\n} e^{a}_{\m} 
\nonumber\\
\6_{\n}e^{a (5)}_{\m 1NS} =
(\ne \nx \6_{\n})e^{a}_{\m} +
\6_{\n}(\ne \nx e^{a}_{\m})
\nonumber\\
\6_{\n}\psi^{\a (0)}_{\m 1} = \6_{\n} \psi^{\a}_{\m} 
\nonumber\\
\6_{\n}\psi^{\a (5)}_{\m 1NS} =
(\nse\nsx \6_{\n} )\psi^{\a}_{\m} + \6_{\n}(\nse\nsx \psi^{\a}_{\m})
\nonumber\\
e^{(0)}_{b\m 1}= e^{(0)}_{b\m 1}
\nonumber\\
e^{(5)}_{b\m 1NS}= \ne\nx e_{b \m}
\nonumber\\
\psi^{\g (0)}_{\m 1} = \nse\nsx \psi^{\g}_{\m}
\nonumber\\
\psi^{\g (5)}_{\m 1NS} = \nse\nsx \psi^{\g}_{\m}
\nonumber\\
(\bar{\e} \g^a \psi_{\m})^{(0)}_{1} = (\bar{\e} \g^a \psi_{\m})
\nonumber\\
(\bar{\e} \g^a \psi_{\m})^{(5)}_{1NS} =
[(\nse\nsx \bar{\e})\g^a \psi_{\m} +
\bar{\e}\g^a (\nse\nsx \psi_{\m})] 
\nonumber\\
(\DD_{\m}\e^{\a (0)})_{1}= \DD_{\m}\e^{\a }
\nonumber\\
(\DD_{\m}\e^{\a (5)})_{1NS}=
(\ne\nx \6_{\m} )\e^{\a} +
\6_{\m}(\nse\nsx \e^{\a})+
\f{1}{2}[\stackrel{\circ}{\o}^{(5)}_{\m abNS}
[\ss^{ab}]^{\a}_{\b}
\e^{\b} +
\nonumber\\
\stackrel{\circ}{\o}^{(0)}_{\m ab}[\ss^{ab}]^{\a}_{\b}
(\nse\nsx \e^{\b})]+ 
\f{1}{8}[\sum_{(\m ,b,c)}[\bar{\psi}_{\m}\g_b \psi_c ][\ss^{bc}]^{\a}_{\b}
(\nse\nsx \e^{b}) +
\nonumber\\
\sum_{(\m ,b,c)}[\bar{\psi}_{\m}\g_b
(\nse\nsx \psi_c )+
(\nse\nsx \bar{\psi}_{\m})\g_b \psi_c ][\ss^{bc}]^{\a}_{\b}\e^{\b}]
\nonumber\\
\stackrel{\circ}{\o}^{(0)}_{\n bc1}= \stackrel{\circ}{\o}_{\n bc}
\nonumber\\
\stackrel{\circ}{\o}^{(5)}_{\n bc1NS}=
(\ne\nx e^{\m}_{a})\6_{[\n }e_{b\m ]} + \cdots
+\f{1}{2}(\ne\nx e^{\r}_{a})e^{\ss}_{b}\6_{\ss}e_{\r c}e^{c}_{\n} +
\f{1}{2}(\nx\ne e^{\r}_{a})e^{\ss}_{b}\6_{\ss}e_{\r c}e^{c}_{\n} + \cdots
\nonumber\\
K^{(0)}_{\n ab1}= K_{\n ab}
\nonumber\\
K^{(5)}_{\n ab1NS}= 
{\i}{8}\sum_{(\n bc)}[\bar{\psi}_{\n}\g_b (\nse \nsx \psi_c )
+ (\nse \nsx \bar{\psi}_{\n})\g_b \psi_c ],
\eea
where $\6_{[\n}e_{b \m ]}$ refers to antisymmetrization with respect to
the indices $\n$ and $\m$ only and the sums on fermion products
in all the expressions used in this paper mean
\be
\sum_{(\n bc)}[A_{\n}\g_b B_c ] =
A_{\n}\g_b B_c - A_{\n}\g_c B_b + A_{b}\g_{\n} B_c .
\ee

\end{document}